\documentclass[aps,prd,showpacs,onecolumn,preprintnumbers,notitlepage,
nofootinbib,amsfont,amssymb,12pt,singlespacing]{revtex4-1}
\usepackage{amsmath}
\usepackage{bm}
\usepackage{times}
\usepackage{braket}
\usepackage{color}
\usepackage{epsfig}
\usepackage{slashed}
\usepackage{hyperref}

\newcommand{\beq}{\begin{eqnarray}}
\newcommand{\eeq}{\end{eqnarray}}
\newcommand{\non}{\nonumber\\ }

\newcommand{\acp}{A_{\rm CP}}

\def\lsim{ {\ \lower-1.2pt\vbox{\hbox{\rlap{$<$}\lower6pt\vbox{\hbox{$\sim$}
}}}\ } }
\def\gsim{ {\ \lower-1.2pt\vbox{\hbox{\rlap{$>$}\lower6pt\vbox{\hbox{$\sim$}
}}}\ } }

\def \jhep{ J. High Energy Phys.  }

\definecolor{Red}{rgb}{1.,0.,0.}

\definecolor{Blue}{rgb}{0.,0.,1.}

\newcommand{\Gre}[1]{{\color{nicegreen}{#1}}}

\definecolor{nicered}{rgb}{0.7,0.1,0.2}
\definecolor{nicegreen}{rgb}{0.1,0.4,0.2}

\hypersetup{colorlinks,citecolor=nicegreen,linkcolor=nicered}

\begin{document}

\title{ Large non-factorizable contributions in $B \to a_0 a_0$ decays}

\author{Defa~Dou}

\author{Xin~Liu\footnote{Corresponding author}\footnote{Electronic address: {\Gre{liuxin@jsnu.edu.cn}}}}

\author{Jing-Wu~Li\footnote{Electronic address: {\Gre{lijw@jsnu.edu.cn}}}}

\affiliation{\small School of Physics and Electronic Engineering,\\
Jiangsu Normal University, Xuzhou, Jiangsu 221116,
People's Republic of China}


\author{Zhen-Jun~Xiao\footnote{Electronic address: {\Gre{xiaozhenjun@njnu.edu.cn}}}}

\affiliation{\small Department of Physics and Institute of Theoretical
Physics,\\
Nanjing Normal University, Nanjing, Jiangsu 210023,
People's Republic of China}


\date{\today}

\begin{abstract}
We investigate three tree-dominated $B \to a_0 a_0$ decays for the first time in the perturbative QCD(pQCD)
approach at leading order in the standard model, with $a_0$ standing for the light scalar $a_0(980)$ state,
which is assumed as a meson based on the model of conventional two-quark$(q\bar q)$ structure. All the
topologies of the Feynman diagrams such as the non-factorizable spectator ones and the annihilation ones
are calculated in the pQCD approach. It is of great interest to find that, contrary to the known
$B \to \pi \pi$ decays, the $B \to a_0 a_0$ decays are governed by the large non-factorizable
contributions, which give rise to the large $B \to a_0 a_0$ decay rates in the order of $10^{-6} \sim 10^{-5}$,
although the $a_0$ meson has an extremely small vector decay constant $f_{a_0}$. Also observed are large direct
CP-violating asymmetries around $15\%$ and $30\%$ for the $B^0 \to a_0^0 a_0^0$ and $a_0^+ a_0^-$ modes.
These sizable predictions could be easily examined at the running Large Hadron Collider and the near future
Super-B/Belle-II experiments. The future precision measurements combined with these pQCD predictions  might
be helpful to explore the complicated QCD dynamics and the inner structure of the light scalar $a_0$,
as well as to complementarily constrain the unitary angle $\alpha$.
\end{abstract}

\preprint{\footnotesize JSNU/PHY-TH-2015}
\maketitle

%
%

\section{Introduction}

As we know, the nature of the light scalar states such as $a_0(980)$
is not yet well understood at both theoretical and experimental
aspects. Also the identification and the classification of these
light scalars remain as a long-standing puzzle(for latest
review, see, e.g.~\cite{Agashe:2014kda}) to be resolved.
However, it is fortunate
for the people that the light scalars as products in the heavy
flavor meson decays have been detected, for example, $D \to SP, SV$,
$B \to SP, SV$, even $B \to SS$
modes~\cite{Agashe:2014kda,Amhis:2014hma} with $S, P,$ and $V$ being the light scalar, pseudoscalar, and vector mesons, respectively, which will provide
unique places and play very important roles on investigating
the physical properties of light scalars. It is generally believed
that the ongoing Large Hadron Collider(LHC) experiments can provide
rich data on the $B$, $B_s$, and $B_c$ meson decaying into light
scalars. And more promisingly, the forthcoming Super-B/Belle-II factory scheduled in 2018 with a high luminosity $\gsim 10^{36} {\rm cm}^{-2} {\rm s}^{-1}$~\cite{Bona:2007qt,Gershon:2006mt} will produce much more events about the
relevant decays. The studies on the above mentioned decays can also
provide more constraints complementarily on the parameters in the standard model(SM), hint the exotic new physics beyond the
SM, etc.

In this work, we will investigate the CP-averaged branching ratios and the CP-violating asymmetries of the $B \to a_0(980) a_0(980)$ decays by employing the perturbative QCD(pQCD) approach~\cite{Keum:2000ph,Lu:2000em,Li:2003yj} with the low energy
effective Hamiltonian~\cite{Buchalla:1995vs} in the SM. It should be
noted that the $a_0(980)$ state here will be assumed as a meson in
the model of conventional two-quark$(q\bar q)$ structure. Moreover,
hereafter, the $a_0(980)$ will be abbreviated as $a_0$ for the sake
of simplicity throughout the paper. To our knowledge, heretofore, no
other $B \to SS$ processes have been studied explicitly in the factorization approaches based on the QCD dynamics, apart from the $B_{u,d,s} \to K_0^*(1430) \bar K_0^*(1430)$ decays~\cite{Liu:2013lka}
by two of our authors (X.~Liu and Z.J.~Xiao). Because the scalar
meson has either tiny or vanishing vector decay
constant~\cite{Cheng:2005nb,Cheng:2009xz}, the
contributions arising from the factorizable emission
diagrams in the $B \to SS$ decays are usually highly
suppressed, which is dramatically different from the
known $B \to PP, PV, VV$ decays. In other words, for
example, in contrast to the extensively investigated
$B \to \pi \pi$ decays, the large measured $B \to a_0 a_0$
decay rates may indicate
large non-factorizable spectator scattering and/or
annihilation contributions, which would
hint some useful information on
the $B \to \pi \pi$ decays, the presently known puzzle to be resolved, because they embrace the
same components at the quark level. In the heavy $B$ meson decays, the above mentioned large
contributions from non-factorizable spectator and annihilation diagrams are often considered
as the small\footnote{In fact, the cancelation of the decay amplitudes indeed occurred between
the two non-factorizable spectator diagrams in the $B \to PP, PV, VV$ channels, for example,
see Ref.~\cite{Liu:2015sra}.} and/or negligible higher order or higher power corrections in
the naive factorization approach~\cite{Bauer:1986bm}. Therefore, the channels involving an
emitted scalar state in the heavy flavor meson decays are suggested to test the breaking
effects of the factorization assumption, e.g.~\cite{Diehl:2001xe}. Though the QCD improved
factorization approach~\cite{Beneke:1999br,Du:2000ff} going beyond the naive factorization,
the end-point singularities make it less predictive because the non-factorizable spectator
scattering contributions and the annihilation ones have to be parametrized with the tunable
parameters, which are always determined by the experimental measurements. As one of the
popular factorization approaches based on the QCD dynamics, the pQCD approach involves no
end-point singularities by retaining the parton transverse momentum $k_T$.
Based on $k_T$ factorization theorem, the double logarithms
arising  from the overlap of soft and collinear divergences
generated in the radiative corrections are resummed into
an important Sudakov factor to suppress the long-distance
contribution~\cite{Li:1996gi}. Armed with this pQCD approach,
all the transition form factors, the non-factorizable
spectator diagrams, and the annihilation diagrams are
perturbatively calculable, besides
the factorizable spectator diagrams. Note that, as far
as the annihilation contributions are concerned, soft-collinear effective theory~\cite{Bauer:2004tj} and pQCD approach have an extremely different effect on the perturbative calculations~\cite{Arnesen:2006vb,Chay:2007ep}. However, the predictions on the pure annihilation decays based on the pQCD approach can accommodate the experimental data well, for example, see Refs.~\cite{Lu:2002iv,Li:2004ep,Ali:2007ff,Xiao:2011tx}.
We will therefore put the controversies aside and adopt the pQCD approach in our analyses.

The paper is organized as follows. Section~\ref{sec:form} is devoted to the analytic expressions for the decay amplitudes of $B \to a_0 a_0$ modes in the pQCD approach. The numerical results and phenomenological analyses on the CP-averaged branching ratios and the CP-violating asymmetries of the considered decays
are given in Sec.~\ref{sec:r&d}. We summarize and conclude in Sec.~\ref{sec:summary}.

%
%
\section{ Perturbative calculations}\label{sec:form}

For the considered $B \to a_0 a_0$ decays,
the related weak effective
Hamiltonian $H_{{\rm eff}}$~\cite{Buchalla:1995vs} can be written as
\beq
H_{\rm eff}\, &=&\, {G_F\over\sqrt{2}}
\biggl\{ V^*_{ub}V_{ud} [ C_1(\mu)O_1^{u}(\mu)
+C_2(\mu)O_2^{u}(\mu) ] 
 - V^*_{tb}V_{td} [ \sum_{i=3}^{10}C_i(\mu)O_i(\mu) ] \biggr\}+ {\rm H.c.}\;,
\label{eq:heff}
\eeq
with the Fermi constant $G_F=1.16639\times 10^{-5}{\rm
GeV}^{-2}$, the Cabibbo-Kobayashi-Maskawa(CKM) matrix elements $V$,
and the Wilson coefficients $C_i(\mu)$ at the renormalization scale
$\mu$. The local four-quark
operators $O_i(i=1,\cdots,10)$ are written as
\begin{enumerate}
\item[]{(1) current-current(tree) operators}
\begin{eqnarray}
{\renewcommand\arraystretch{1.5}
\begin{array}{ll}
\displaystyle
O_1^{u}\, =\,
(\bar{d}_\alpha u_\beta)_{V-A}(\bar{u}_\beta b_\alpha)_{V-A}\;,
& \displaystyle
O_2^{u}\, =\, (\bar{d}_\alpha u_\alpha)_{V-A}(\bar{u}_\beta b_\beta)_{V-A}\;;
\end{array}}
\label{eq:operators-1}
\end{eqnarray}

\item[]{(2) QCD penguin operators}
\begin{eqnarray}
{\renewcommand\arraystretch{1.5}
\begin{array}{ll}
\displaystyle
O_3\, =\, (\bar{d}_\alpha b_\alpha)_{V-A}\sum_{q'}(\bar{q}'_\beta q'_\beta)_{V-A}\;,
& \displaystyle
O_4\, =\, (\bar{d}_\alpha b_\beta)_{V-A}\sum_{q'}(\bar{q}'_\beta q'_\alpha)_{V-A}\;,
\\
\displaystyle
O_5\, =\, (\bar{d}_\alpha b_\alpha)_{V-A}\sum_{q'}(\bar{q}'_\beta q'_\beta)_{V+A}\;,
& \displaystyle
O_6\, =\, (\bar{d}_\alpha b_\beta)_{V-A}\sum_{q'}(\bar{q}'_\beta q'_\alpha)_{V+A}\;;
\end{array}}
\label{eq:operators-2}
\end{eqnarray}

\item[]{(3) electroweak penguin operators}
\begin{eqnarray}
{\renewcommand\arraystretch{1.5}
\begin{array}{ll}
\displaystyle
O_7\, =\,
\frac{3}{2}(\bar{d}_\alpha b_\alpha)_{V-A}\sum_{q'}e_{q'}(\bar{q}'_\beta q'_\beta)_{V+A}\;,
& \displaystyle
O_8\, =\,
\frac{3}{2}(\bar{d}_\alpha b_\beta)_{V-A}\sum_{q'}e_{q'}(\bar{q}'_\beta q'_\alpha)_{V+A}\;,
\\
\displaystyle
O_9\, =\,
\frac{3}{2}(\bar{d}_\alpha b_\alpha)_{V-A}\sum_{q'}e_{q'}(\bar{q}'_\beta q'_\beta)_{V-A}\;,
& \displaystyle
O_{10}\, =\,
\frac{3}{2}(\bar{d}_\alpha b_\beta)_{V-A}\sum_{q'}e_{q'}(\bar{q}'_\beta q'_\alpha)_{V-A}\;.
\end{array}}
\label{eq:operators-3}
\end{eqnarray}
\end{enumerate}
with the color indices $\alpha, \ \beta$ and the notations
$(\bar{q}'q')_{V\pm A} = \bar q' \gamma_\mu (1\pm \gamma_5)q'$.
The index $q'$ in the summation of the above operators runs
through $u,\;d,\;s$, $c$, and $b$.
The standard combinations $a_i$ of Wilson coefficients are defined as follows,
  \beq
a_1&=& C_2 + \frac{C_1}{3}\;, \qquad  a_2 = C_1 + \frac{C_2}{3}\;,\quad
 a_i = C_i + \frac{C_{i \pm 1}}{3}(i=3 - 10) \;.
  \eeq
where the upper(lower) sign applies, when $i$ is odd(even).

\begin{figure}[!!hbt]
  \centering
  \begin{tabular}{c}
  \includegraphics[width=0.85\textwidth]{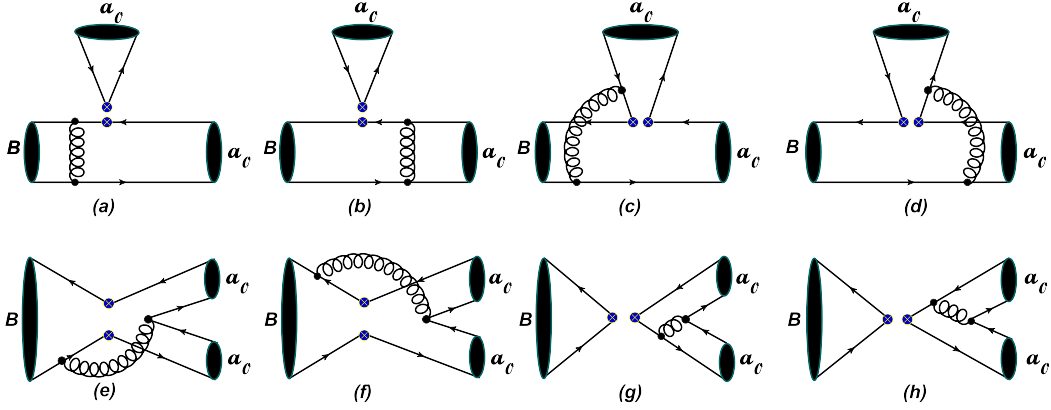}
  \end{tabular}
  \caption{(Color online) Typical Feynman diagrams for $B \to a_0
a_0$ decays at leading order in the pQCD approach.}
  \label{fig:fig1}
\end{figure}

Similar to $B \to \pi \pi$ decays~\cite{Lu:2000em,Li:2005kt},
there are eight types of diagrams contributing to $B \to a_0 a_0$
modes at leading order(LO) in the pQCD approach, as illustrated
in Fig.~\ref{fig:fig1}. They can be classified into two types
of topologies as emission 
and annihilation, respectively. And
each kind of topology contains factorizable diagrams such
as  Fig.~\ref{fig:fig1}(a) and~\ref{fig:fig1}(b), in which a
hard gluon connects the quarks in the same meson, and
non-factorizable diagrams such as Fig.~\ref{fig:fig1}(c)
and~\ref{fig:fig1}(d), in which a hard gluon attaches the
quarks in two different mesons. By evaluating all these
Feynman diagrams, one can obtain the decay amplitudes of
$B \to a_0 a_0$ decays. Because the above mentioned diagrams
are the same as those in $B \to K_0^*(1430) \bar K_0^*(1430)$
modes~\cite{Liu:2013lka}, and also the light scalar mesons
are considered, the formulas of $B \to a_0 a_0$ decays are
therefore same as those of $B \to K_0^*(1430) \bar K_0^*(1430)$
ones just by replacing the wave functions and input
parameters correspondingly. Hence 
the analytic formulas for
the $B \to a_0 a_0$ decays are not 
explicitly presented
in this paper.

By taking various 
contributions from the relevant Feynman
diagrams into consideration, the total decay amplitudes for
three tree-dominated $B \to a_0 a_0$ channels can then be read as,

\begin{enumerate}

\item for $B^0 \to a_0^+ a_0^-$ decay mode,
\begin{widetext}
\beq
{\cal A}(B^0 \to a_0^+ a_0^-) &=& \lambda_u \biggl[  C_1 M_{nfs} + C_2 M_{nfa}  \biggr] - \lambda_t \biggl[   (C_3 + C_9) M_{nfs}  + (C_3 + 2 C_4 -\frac{1}{2}
(C_9\non  & & - C_{10}))  M_{nfa}  + (C_5
- \frac{1}{2} C_7) M_{nfa}^{P_1} + (2 C_6 + \frac{1}{2} C_8) M_{nfa}^{P_2}  + (a_6
- \frac{1}{2} a_8)\non  & & \times f_B F_{fa}^{P_2} + (a_6 +a_8) F_{fs}^{P2}   \biggr]\;, \label{eq:tda-b02a0pa0m}
\eeq
\end{widetext}
where  $\lambda_{u}= V_{ub}^* V_{ud}$ and $\lambda_t = V_{tb}^* V_{td}$. We adopt $F$ and
$M$ to denote the contributions from $(V-A)(V-A)$ operators in the factorizable and non-factorizable diagrams, respectively. Analogously, $F^{P1}$ and
$M^{P1}$ are chosen to denote the contributions from $(V-A)(V+A)$
operators, and $F^{P2}$ and $M^{P2}$ are taken to denote the contributions from $(S-P)(S+P)$ operators which result from the Fierz transformation of the $(V-A)(V+A)$ operators. The subscripts $fs$, $nfs$, $fa$, and $nfa$ are the abbreviations for
factorizable emission, 
non-factorizable emission, 
factorizable annihilation, and non-factorizable annihilation, respectively.

\item for $B^+ \to a_0^+ a_0^0$ decay mode,
  \begin{widetext}
\beq 
\sqrt{2} {\cal A}(B^+ \to a_0^+ a_0^0) &=& \lambda_u \biggl[(C_1 + C_2) M_{nfs} \biggr] - \lambda_t \biggl[
  \frac{1}{2} (C_7 +3 C_8) F_{fs}^{P2}
+ \frac{3}{2} (C_9 + C_{10}) M_{nfs} \non &&   +
 \frac{3}{2} C_8 M_{nfs}^{P_2} \biggr]\;,
\label{eq:tda-bp2a0pa00}
\eeq
\end{widetext}

\item for $B^0 \to a_0^0 a_0^0$ decay mode,
 \begin{widetext}
\beq 
\sqrt{2} {\cal A}(B^0 \to a_0^0 a_0^0) &=& \lambda_u \biggl[ C_2 (M_{nfa} - M_{nfs})\biggr] -\lambda_t \biggl[ -(a_6 - \frac{1}{2} a_8) F_{fs}^{P2}
 + (C_3 - \frac{1}{2}
(C_9\non &&  + 3 C_{10})) M_{nfs} - \frac{3}{2}
C_8 M_{nfs}^{P_2}  + (C_3+ 2 C_4 - \frac{1}{2} (C_9 - C_{10})) M_{nfa} \non &&  +
(C_5 - \frac{1}{2} C_7) M_{nfa}^{P_1} + (2 C_6 + \frac{1}{2} C_8) M_{nfa}^{P_2} + (a_6
- \frac{1}{2} a_8) f_B F_{fa}^{P_2}\biggr]. \label{eq:tda-b02a00a00}
\eeq
\end{widetext}
\end{enumerate}
It is worth mentioning that the highly suppressed $F_{fs}$ has
been safely neglected in all of the above decay amplitudes for
the considered $B \to a_0 a_0$ decays due to the either extremely
small or vanishing vector decay constant.  Furthermore, based on the discussions of $F_{fa}$ below Eq.~(40) in Ref.~\cite{Liu:2013lka}, the factorizable annihilation contributions induced by the $V\pm A$ currents are therefore
naturally absent because of the isospin symmetry between $u$ and $d$ quarks in the above analytical decay amplitudes.

\section{Numerical Results and Discussions} \label{sec:r&d}

In this section, we will make 
theoretical predictions on the
CP-averaged branching ratios and the CP-violating asymmetries for the
 $B \to a_0 a_0$ decay modes considered.
In numerical calculations, central values of the input parameters will be
used implicitly unless otherwise stated. Firstly, we shall make several essential discussions on the input quantities.

\subsection{Input quantities}

For $B$ meson, the distribution amplitude in the impact $b$
space, with $b$ being the conjugate space coordinate of transverse momentum $k_T$, has been proposed~\cite{Keum:2000ph,Lu:2000em,Li:2003yj},
\beq
\phi_{B}(x,b)&=& N_Bx^2(1-x)^2
\exp\left[-\frac{1}{2}\left(\frac{xm_B}{\omega_b}\right)^2
-\frac{\omega_b^2 b^2}{2}\right] \;,
\eeq
where the normalization factor $N_{B}$
is related to the decay constant $f_{B}$ through the following normalization condition,
\beq
\int_0^1 dx \phi_{B}(x, b=0) &=& \frac{f_{B}}{2 \sqrt{2N_c}}\;.\label{eq:norm}
\eeq
with the color factor $N_c=3$. The shape parameter $\omega_b$ has been fixed at $0.40$~GeV associated with $N_B =91.745$ by using the rich experimental
data on the $B$ mesons with $f_{B}= 0.19$~GeV based on lots of calculations of form factors
and other well-known decay modes of $B$ meson
in the pQCD approach~\cite{Lu:2000em,Keum:2000ph,Lu:2002ny}.

For the light scalar $a_0$, its leading twist light-cone distribution amplitude $\phi_{a_0}(x,\mu)$ can be generally expanded as the Gegenbauer polynomials~\cite{Cheng:2005nb,Li:2008tk}:
\beq
\phi_{a_0}(x,\mu)&=&\frac{3}{\sqrt{2N_c}}x(1-x)\biggl\{f_{a_0}(\mu)+\bar
f_{a_0}(\mu)\sum_{m=1}^\infty B_m(\mu)C^{3/2}_m(2x-1)\biggr\}, \label{eq:a0-twist-2}
\eeq
where $f_{a_0}(\mu)$ and $\bar f_{a_0}(\mu)$, $B_m(\mu)$, and
$C_m^{3/2}(t)$ are the vector and scalar decay constants,
Gegenbauer moments, and Gegenbauer polynomials,
respectively. For the vector and scalar decay constants, $
 \bar f_{a_0} = \mu_{a_0} f_{a_0}$ with $ \mu_{a_0} = \frac{m_{a_0}}{m_2(\mu)-m_1(\mu)}$ and $m_{a_0}=0.985$ GeV,
where $m_1$ and $m_2$ are the running current quark masses in the scalar $a_0$.
For neutral scalar $a_0^0$ meson, which cannot be produced by the vector current, the vector decay constant $f_{a_0^0}=0$ is
guaranteed by charge conjugation invariance.
But the quantity $\bar f_{a_0}=f_{a_0} \mu_{a_0}$ remains finite.
In fact, for the charged $a_0^\pm$ meson, the vector decay constant
$f_{a_0^\pm}$ also vanishes in the isospin limit. The reason is
that $f_{a_0^\pm}$ is proportional to the mass difference between the constituent $d$ and $u$ quarks, which will result in $f_{a_0^\pm}$ being of order $m_d -m_u$.
Hence, the contribution from the first term in Eq.~(\ref{eq:a0-twist-2}), namely, $f_{a_0}$, can be neglected
safely.
In other words, the factorizable spectator diagrams could not contribute to $B \to a_0 a_0$
decays through the vector currents. We shall use the same
light-cone distribution amplitudes for both neutral and charged $a_0$ mesons for simplicity in this paper.

The values for scalar decay constant and Gegenbauer moments in
the $a_0$ distribution amplitudes have been investigated at scale $\mu=1~
\mbox{GeV}$~\cite{Cheng:2005nb}:
\beq
\bar f_{a_0}&=& 0.365 \pm 0.020~{\rm GeV}, \quad
B_1=-0.93 \pm 0.10\;, \quad  B_3 = 0.14 \pm 0.08\;.
\label{eq:a0-para}
\eeq

As for the twist-3 distribution amplitudes $\phi_{a_0}^S$ and
$\phi_{a_0}^T$, we here adopt the asymptotic forms
in our numerical calculations for simplicity~\cite{Cheng:2005nb}:
\beq
\phi^S_{a_0}&=& \frac{1}{2\sqrt {2N_c}}\bar f_{a_0},\quad
\phi_{a_0}^T=
\frac{1}{2\sqrt {2N_c}}\bar f_{a_0}(1-2x).
\eeq

The QCD scale~({\rm GeV}), masses~({\rm GeV}),
 and $B$ meson lifetime({\rm ps}) are~\cite{Keum:2000ph,Lu:2000em,Agashe:2014kda}
\beq
 \Lambda_{\overline{\mathrm{MS}}}^{(f=4)} &=& 0.250\; , \quad m_W = 80.41\;,
 \quad  m_{B}= 5.2792\;, \quad m_b = 4.8 \;; \non
  \tau_{B^+} &=& 1.643\;,  \quad \tau_{B^0}= 1.53\;,
   \quad  m_{a_0} =0.985\;.
\label{eq:mass}
\eeq

For the CKM matrix elements, we adopt the Wolfenstein
parametrization and the updated parameters $A=0.814$,
 $\lambda=0.22537$, $\bar{\rho}=0.117 \pm 0.021$, and $\bar{\eta}=0.353 \pm 0.013$~\cite{Agashe:2014kda}.

Utilizing the above chosen distribution amplitudes and the relevant
input parameters, we can get the numerical results in the pQCD approach for the form
factor $F_{0,1}^{B \to a_0}$~\footnote{ The form factor $F_{0,1}^{B \to a_0}$ can
be extracted directly from Eq.~(29) in~\cite{Liu:2013lka} with the state $S$ being $a_0$.
Of course, the readers can also refer to Ref.~\cite{Li:2008tk} for more details.} at maximal recoil as follows,
  \beq
F_{0,1}^{B \to a_0} (q^2=0)
&=& 0.40^{+0.05}_{-0.06}(\omega_{b})^{+0.02}_{-0.02}(\bar f_{a_0})^{+0.02}_{-0.02}(B_{i}^{a_0})   \label{eq:formf-b}  \;,
  \eeq
where the errors arise from the shape parameter $\omega_b$ in $B$
meson distribution amplitude, the scalar decay constant $\bar f_{a_0}$, and the Gegenbauer moments
$B_i^{a_0}(i=1,3)$ in the light $a_0$ distribution amplitude, respectively.
This value agrees well with $0.39^{+0.10}_{-0.08}$ as given in Ref.~\cite{Li:2008tk}.  The tiny deviation is just from the zero
vector decay constant $f_{a_0}$ assumed in this work.

\subsection{CP-averaged branching ratios and CP-violating asymmetries}\label{ssec:cp-brs}

In this subsection, we will analyze the CP-averaged $B \to a_0 a_0$ branching ratios and the CP-violating asymmetries in the pQCD approach at LO level.
For $B \to a_0 a_0$ decays, the decay rate can be written as
\beq
\Gamma =\frac{G_{F}^{2}m^{3}_{B}}{32 \pi  } (1-2 r_{a_0}^2) |{\cal A}(B
\to a_0 a_0)|^2\;,\label{eq:bqdr}
\eeq
where the decay amplitudes ${\cal A}$  can be referred correspondingly in Eqs.~(\ref{eq:tda-b02a0pa0m}-\ref{eq:tda-b02a00a00}).
Using the decay amplitudes
obtained in last section, it is straightforward to numerically evaluate the CP-averaged branching ratios with errors as collected in
Eqs.~(\ref{eq:bra98pm})-(\ref{eq:bra98zz}),
\beq
Br(B^0 \to {a_0}^+ {a_0}^-) &=&
1.5^{+0.7}_{-0.5}(\omega_{b})^{+0.3}_{-0.3}(\bar f_{a_0})
^{+0.7}_{-0.6}(B_{i}^{a_0})
^{+0.1}_{-0.1}({\rm CKM})
\times  10^{-5}  \label{eq:bra98pm}  \;,\\
Br(B^+ \to {a_0}^+ {a_0}^0) &=&
6.1^{+2.6}_{-2.1}(\omega_b)^{+1.4}_{-1.2}(\bar f_{a_0})
^{+3.1}_{-2.2}(B_{i}^{a_0})
^{+0.4}_{-0.4}({\rm CKM})
 \times  10^{-6}  \label{eq:bra98pz}  \;, \\
Br(B^0 \to {a_0}^0 {a_0}^0)  &=&
2.7^{+1.1}_{-1.0}(\omega_{b})^{+0.6}_{-0.6}(\bar f_{a_0})
^{+1.3}_{-1.0}(B_{i}^{a_0})
^{+0.1}_{-0.2}({\rm CKM})
\times  10^{-5}  \label{eq:bra98zz} \; ;
\eeq
The dominant errors are induced by the uncertainties of the
shape parameter $\omega_b = 0.40 \pm 0.04$~GeV for $B$ meson,
the scalar decay constant $\bar f_{a_0}$, and
the Gegenbauer moments $B^{a_0}_i(i=1, 3)$ for the scalar $a_0$(see Eq.~(\ref{eq:a0-para}) for detail), respectively.
It is worth stressing that the effective constraints on the above mentioned non-perturbative parameters 
might be helpful to explore the QCD dynamics involved in these decays and to reveal the inner structure of the light
scalar $a_0$ state.

From Eqs.~(\ref{eq:bra98pm})-(\ref{eq:bra98zz}), one can obviously
observe that the large $B \to a_0 a_0$ decay rates are in the order
of $10^{-6} \sim 10^{-5}$ calculated in the pQCD approach at LO level,
which could be easily detected  through the dominant $a_0$ to $\eta \pi$(or $\pi \pi )$ final state~\cite{Aubert:2004hs} at the running LHC and the forthcoming Super-B/Belle-II experiments.
As mentioned in the Introduction, some
decays involving scalar mesons were suggested as the ideal channels
to test the validation of the factorization assumption~\cite{Diehl:2001xe}.
It is therefore worth stressing that the $B^+ \to a_0^+ a_0^0$
mode would be the best choice, because it 
only contains a
significantly suppressed factorizable emission 
contribution
and a negligible non-factorizable emission 
contribution as
proposed in naive factorization, but 
has a large branching ratio
that could be easily tested in the near future experiments.
Therefore, the observation of this large
$B^+ \to a_0^+ a_0^0$ decay rate,
on one hand, could offer an effective test to the breaking effects
of the factorization assumption; on the other hand, {\it might} verify
the $q\bar q$ components of the light scalar $a_0$ evidently.
Furthermore, it is surprising to find that the conventionally so-called "color-suppressed"
$B^0 \to a_0^0 a_0^0$ mode has the largest branching ratio as $2.7 \times 10^{-5}$, which
is highly different from the known color-suppressed $B \to PP$ modes, such as the famous
$B^0 \to \pi^0 \pi^0$ channel with very small branching ratio around ${\cal O}(10^{-7})$,
although they embrace the same components at the quark level. Consequently, the hierarchy
of the branching ratios exhibits theoretically as $Br(B^0 \to a_0^0 a_0^0)
\sim Br(B^0 \to a_0^+ a_0^-) > Br(B^+ \to a_0^+ a_0^0)$ in the pQCD approach, which is also
dramatically different from that in the $B \to \pi \pi$ decays as  $Br(B^0 \to \pi^+ \pi^-)
\gtrsim Br(B^+ \to \pi^+ \pi^0) >> Br(B^0 \to \pi^0 \pi^0)$ within theoretical errors~\cite{Lu:2000em,Li:2005kt,Xiao:2011tx,Liu:2015sra} and $Br(B^+ \to \pi^+ \pi^0) \gtrsim Br(B^0 \to \pi^+ \pi^-)> Br(B^0 \to \pi^0 \pi^0)$ within experimental uncertainties~\cite{Agashe:2014kda,Amhis:2014hma}, respectively. In
terms of the central values of the $B \to a_0 a_0$ decay rates, the following relation can be easily found,
\beq
Br(B^0 \to a_0^0 a_0^0) &>& Br(B^0 \to a_0^+ a_0^-)
> Br(B^+ \to a_0^+ a_0^0)\;, \label{eq:re-a0}
\eeq
which can be traced back to the factorization formulas as given in
Eqs.~(\ref{eq:tda-b02a0pa0m})-(\ref{eq:tda-b02a00a00}). Specifically,
the tree dominant contributions of these three decays are
$C_2\ (M_{nfa} - M_{nfs})$, $C_1\ M_{nfs} + C_2\ M_{nfa}$,
and $(C_1 + C_2) M_{nfs}$, respectively, in which $C_2$ is
much larger than $C_1$ in magnitude with $C_2 \sim 1.12$
and $C_1 \sim -0.27$ at the $m_b$ scale, and $M_{nfs}(M_{nfa})$
stands for the amplitude of the non-factorizable emission 
(annihilation)
diagrams induced by the tree operators $O_{1,2}$.
\begin{table*}[b]
\caption{ The factorization decay amplitudes(in unit of $10^{-3}\; \rm{GeV}^3$) of the charmless hadronic $B \to a_0 a_0$
decays in the pQCD approach at leading order level, where only the central values are quoted for clarification.}
\label{tab:DA-a0a0}
\begin{center}\vspace{-0.2cm}
{ \begin{tabular}[t]{l|c|c|c|c}
 \hline \hline
 Decay modes     &    ${\cal A}_{fs}$      &  ${\cal A}_{nfs}$          &  ${\cal A}_{nfa}$              &    ${\cal A}_{fa}$           \\
\hline
$B^0 \to a_0^+ a_0^-$              &$\hspace{0.25cm} 0.950-{\it i} 0.390$   &$\hspace{0.25cm} 1.619-{\it i} 2.982$
                                             &$ -1.056- {\it i} 1.876$  &$ -0.044+ {\it i} 1.212$
\\
$B^+ \to a_0^+ a_0^0$             &$ -0.018+ {\it i} 0.007$  &$ -1.268+ {\it i} 2.926$
                                            &$0.0$  &$ 0.0$
\\
$B^0 \to a_0^0 a_0^0$        &$\hspace{0.25cm} 0.691- {\it i} 0.284$  &$\hspace{0.25cm} 2.458- {\it i} 5.100$
                                             &$-0.799- {\it i} 1.363$  &$-0.035+ {\it i} 0.853$
 \\
\hline \hline
\end{tabular} }
\end{center}
\end{table*}
The underlying reason is that, as presented in Eq.~(\ref{eq:a0-twist-2}),
the asymmetric leading twist distribution amplitude $\phi_{a_0}(x)$
turns the originally destructive interferences induced by the
symmetric one $\phi_{P}^A(x)$ between the two non-factorizable
emission 
diagrams, namely, Fig.~\ref{fig:fig1}(c) and~\ref{fig:fig1}(d), in the $B \to PP$ decays into the presently constructive ones in the $B \to a_0 a_0$ modes. Meanwhile, the analogous phenomenon also occurs in the annihilation topologies. Note that the
values of $M_{nfa}$ are usually a bit smaller than those of
$M_{nfs}$ in modulus, because the former is always power
$1/m_B$ suppressed with $m_B$ being the $B$ meson mass.
It is interesting to note that the QCD behavior in light
scalar $a_0$ is greatly different from that in the
pseudoscalar pion, which can be seen apparently that
the leading twist $a_0$(pion) distribution amplitude
is governed by the odd(even) Gegenbauer
polynomials~\cite{Cheng:2005nb,Chernyak:1983ej,Ball:1998tj}. Therefore,
large non-factorizable contributions are observed in the $B \to a_0 a_0$ decays.

\begin{table*}[htb]
\caption{ Same as Table~\ref{tab:DA-a0a0} but for the charmless hadronic $B \to \pi \pi$ decays.}
\label{tab:DA-pi0pi0}
\begin{center}\vspace{-0.6cm}
{\small
\begin{tabular}[t]{l|c|c|c|c}
 \hline \hline
 Channels     &    ${\cal A}_{fs}$      &  ${\cal A}_{nfs}$ &    ${\cal A}_{nfa}$         &  $ {\cal A}_{fa}$       \\
\hline \hline
$B^0 \to \pi^+ \pi^-$             &$ -1.845-{\it i} 2.957$  &$\hspace{0.25cm} 0.095+{\it i} 0.075$&$ -0.047+ {\it i} 0.159$  &$ 0.038+ {\it i} 0.196$
\\
$B^+ \to \pi^+ \pi^0$             &$-0.844- {\it i} 1.988$  &$ -0.086- {\it i} 0.082$  &$ 0.0$  &$0.0$
\\
$B^0 \to \pi^0 \pi^0$              &$ -0.461- {\it i} 0.104$   &$\hspace{0.25cm} 0.153+ {\it i} 0.135$&$ -0.033+ {\it i} 0.113$  &$  0.029+ {\it i} 0.139$
 \\
  \hline \hline
 \end{tabular} }
\end{center}
\end{table*}


In view of the surprisingly large $Br(B^0 \to a_0^0 a_0^0)$ and the amazingly small $Br(B^0 \to \pi^0 \pi^0)$ in the pQCD approach at LO level, respectively,
we here present the numerical decay amplitudes\footnote{The topological
amplitudes ${\cal A}_{fs}, {\cal A}_{nfs}, {\cal A}_{nfa}$,
and ${\cal A}_{fa}$ shown in the Tables~\ref{tab:DA-a0a0}
and~\ref{tab:DA-pi0pi0} stand for the decay amplitudes
of factorizable emission, 
non-factorizable emission, 
non-factorizable annihilation, and factorizable
annihilation diagrams, respectively.}(See Tables~\ref{tab:DA-a0a0}
and~\ref{tab:DA-pi0pi0} for detail) arising from every topology
to clarify the aforementioned predictions explicitly. It can be
clearly seen that the decay amplitudes in the $B \to a_0 a_0$
decays exhibit very different pattern from those in the
$B \to \pi \pi$ ones, although they embrace the same
diagrams at the quark level: the former modes determined
by the non-factorizable contributions with a larger scalar
decay constant $\bar f_{a_0} \sim 0.365$ GeV, while the
latter ones dominated by the factorizable emission 
contributions with a smaller $f_\pi \sim 0.130$ GeV, apart from the special $B^0 \to \pi^0 \pi^0$ channel. As mentioned above, the underlying reason is that these considered modes include dramatically different QCD dynamics. Notice that, for the $B \to a_0 a_0$ decays, because of the vanished vector decay constant $f_{a_0} \sim 0$, ${\cal A}_{fs}$ come only from the penguin contributions induced by the $(S+P)(S-P)$ operators, which are from the $(V+A)(V-A)$ ones by Fierz transformation. However, the phenomenologies shown in
$B \to a_0 a_0$ decays indicate that the famous $B \to \pi \pi$ puzzle could be resolved if a new QCD mechanism is resorted to enhance the non-factorizable contributions. Of course, it is nontrivial to resolve the $B \to \pi \pi$ puzzle just by including the large non-factorizable
contributions. This point has been clarified in the literatures, for example, see Refs.~\cite{Li:2009wba,Liu:2015sra}.

Because of the large errors induced by the much less constrained hadronic parameters such as the scalar decay constant $\bar f_{a_0}$,
the Gegenbauer moments $B_1$ and $B_3$ in the $a_0$ distribution amplitudes, we derive the ratios of the branching ratios, in which the parameter uncertainties may be greatly canceled and be more helpful for measurements in the relevant experiments,
\beq
R_{0+} &\equiv& \frac{Br(B^0 \to a_0^+ a_0^-)}{Br(B^+ \to a_0^+ a_0^0)} \approx
2.44^{+0.06}_{-0.01}(\omega_b)^{+0.00}_{-0.00}(\bar f_{a_0})
^{+0.06}_{-0.01}(B_i^{a_0})^{+0.01}_{-0.01}({\rm CKM})\;, \\
R_{00} &\equiv& \frac{Br(B^0 \to a_0^+ a_0^-)}{Br(B^0 \to a_0^0 a_0^0)} \approx
0.56^{+0.01}_{-0.00}(\omega_b)^{+0.00}_{-0.00}(\bar f_{a_0})
^{+0.01}_{-0.00}(B_i^{a_0})^{+0.00}_{-0.00}({\rm CKM})\;, \\
R_{+0} &\equiv& \frac{Br(B^+ \to a_0^+ a_0^0)}{Br(B^0 \to a_0^0 a_0^0)} \approx
0.23^{+0.00}_{-0.00}(\omega_b)^{+0.00}_{-0.00}(\bar f_{a_0})
^{+0.00}_{-0.00}(B_i^{a_0})^{+0.00}_{-0.00}({\rm CKM})\;;
\eeq

\begin{figure}[!!hbt]
  \centering
  \begin{tabular}{lll}
  \includegraphics[height=5.0cm,width=5.5cm]{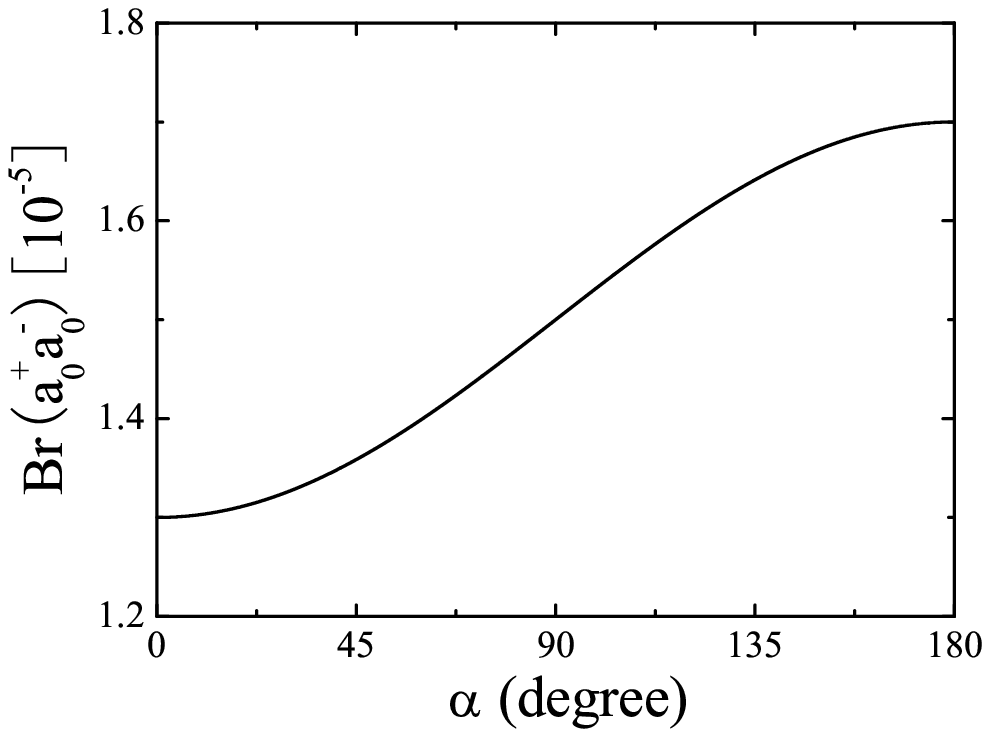}
 &\includegraphics[height=5.0cm,width=5.5cm]{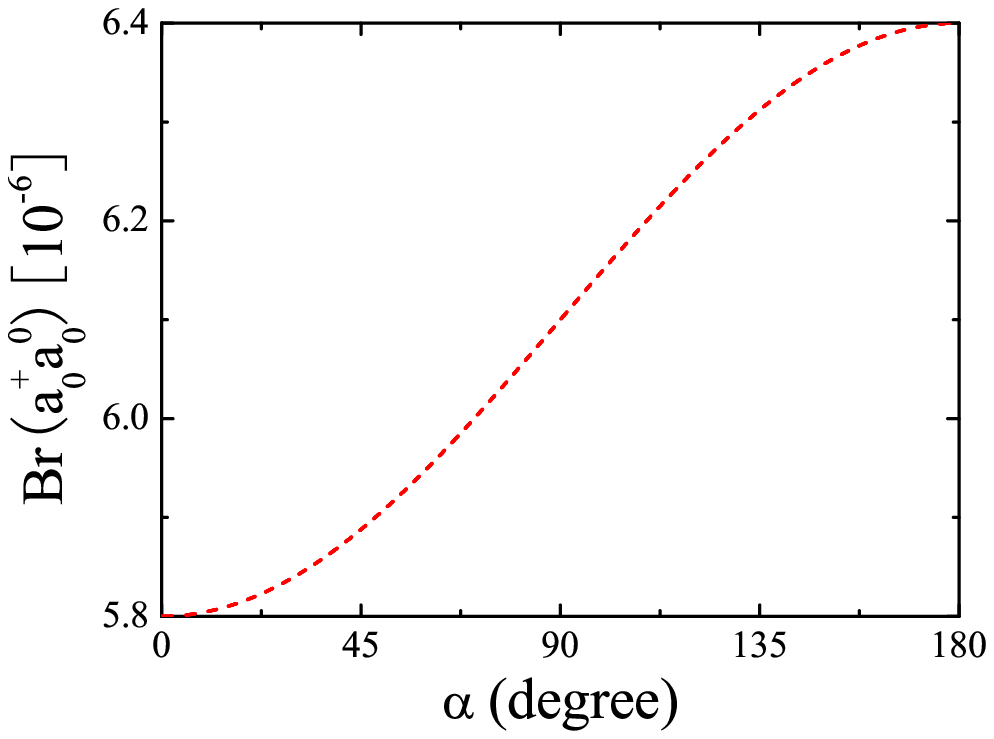}
 &\includegraphics[height=5.0cm,width=5.5cm]{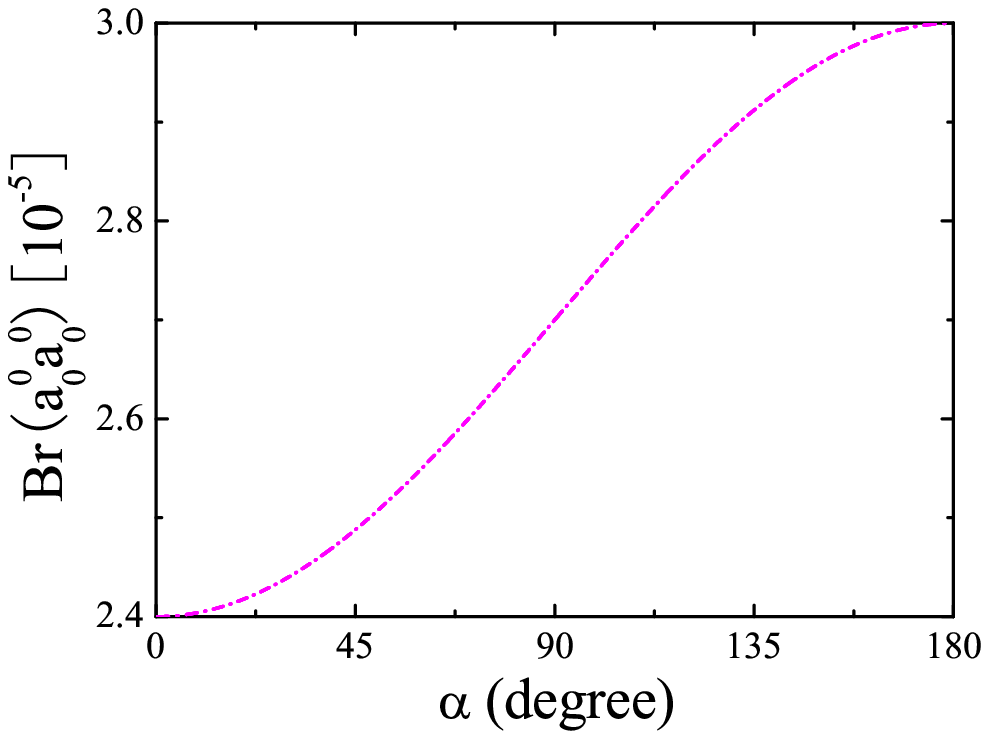}
  \end{tabular}
  \caption{(Color online) Dependence on the CKM angle $\alpha$ of
  the $B \to a_0^+ a_0^-$(Solid line), $a_0^+ a_0^0$(Dashed line),
  and $a_0^0 a_0^0$(Dash-dotted line) decay rates at leading order
  in the pQCD approach, respectively.}
  \label{fig:fig2}
\end{figure}

It is well known that the $B \to \pi \pi$ modes can provide
important information to constrain the CKM unitary angle $\alpha$.
As they contain the same quark diagrams as the
$B \to \pi \pi$ decays, it is generally believed
that the $B \to a_0 a_0$ processes can also
provide complementary constraints on the angle $\alpha$. Here, we show the $\alpha$ dependent branching ratios of the $B \to a_0 a_0$ decays in the pQCD approach at the LO level. Based on Eqs.~(\ref{eq:tda-b02a0pa0m})-(\ref{eq:tda-b02a00a00}), the decay amplitudes of $B\to a_0 a_0$ decays can be rewritten as follows,
\beq
{\cal A}&=& V_{ub}^*V_{ud}T-V_{tb}^*V_{td}P
=V_{ub}^*V_{ud}T(1+ze^{i(\alpha+\delta)})\;,
\label{eq:DA-B}
\eeq
where the weak phase $\alpha=\arg \left[-\frac{V_{tb}^*V_{td}}{V_{ub}^*V_{ud}}\right]$, the ratio  $z=|V_{tb}^*V_{td}/V_{ub}^*V_{ud}| \cdot |P/T|$, and $\delta$ is the
relative strong phase between tree($T$) and penguin($P$) amplitudes. Correspondingly, the decay amplitudes of the $\bar B \to a_0 a_0$ decays can be read as,
\beq
\overline{\cal A} &=& V_{ub}V_{ud}^*T-V_{tb}V_{td}^*P
=V_{ub}V_{ud}^*T(1+ze^{i(-\alpha+\delta)})\;,
\label{eq:DA-Bb}
\eeq
Therefore, the CP-averaged branching ratio of the $B \to a_0 a_0$ decays shall be the following,
\beq
Br(B \to a_0 a_0) &=&(|{\cal A}|^2+|\overline{{\cal A}}|^2)/2=|V_{ub}^*V_{ud}T|^2(1+2z\cos\alpha\cos\delta+z^2)\;.
\label{eq:Br-cP}
\eeq
It is thus easy to see that the CP-averaged
branching ratio is a function of $\cos\alpha$ for the given ratio $z$
and the strong phase $\delta$, which can be perturbatively calculated in the pQCD approach. This gives a potential method to
determine the CKM angle $\alpha$ by measuring the
CP-averaged branching ratios with precision. The dependence on
the CKM weak phase $\alpha$ of the CP-averaged branching ratios
for $B^0 \to a_0^+ a_0^-$(Solid line), $B^+ \to a_0^+ a_0^0$(Dashed
line), and $B^0 \to a_0^0 a_0^0$(Dash-dotted line) decays,
respectively, are presented in Fig.~\ref{fig:fig2}, where
the central values of the predictions in the pQCD approach
are simply quoted for clarification. Then we can directly
observe that the central decay rates for the $B \to a_0 a_0$
decays in the pQCD approach at LO level correspond to the
value around $90^\circ$ of the CKM angle $\alpha$, which
agrees well with the constraints from various
experiments~\cite{Agashe:2014kda}.


Now we turn to the evaluations of the CP-violating asymmetries of $B
\to a_0 a_0$ decays in the pQCD approach.
For $B^+ \to a_0^+ a_0^0$ decay, the direct CP-violating
asymmetry $\acp$ can be defined as:
 \beq
\acp^{\rm dir} =  \frac{|\overline{\cal A}_f|^2 - |{\cal A}_f|^2}{
 |\overline{\cal A}_f|^2+|{\cal A}_f|^2},
\label{eq:acp1}
\eeq
Using Eq.~(\ref{eq:acp1}), it is easy to calculate the
direct CP-violating asymmetry for the considered
 $B^+ \to a_0^+ a_0^0$ mode as listed in
 Eq.~(\ref{eq:adir98}),
\beq
\acp^{\rm dir}(B^+ \to a_0^+ a_0^0) & = &
-0.6^{+0.1}_{-0.2}(\omega_{b})^{+0.1}_{-0.2}(B_{i}^{a_0})
^{+0.0}_{-0.1}({\rm CKM})  \%  \label{eq:adir98} \; ,
\eeq
This tiny direct CP-violating asymmetry would be hard to be
measured because of the extremely small penguin contributions
in magnitude, although the large strong phase can be obtained
due to the constructive interferences between the two
non-factorizable emission 
diagrams with the asymmetric $a_0$ leading twist distribution
amplitude, which is very different from that in the
$B^+ \to \pi^+ \pi^0$ mode with the small non-factorizable emission
contributions, relative to the
purely real amplitudes from the factorizable emission diagrams in the pQCD approach at LO level.

As to the CP-violating asymmetries for the neutral decays $B^0 \to
a_0 a_0$, the effects of $B^0-\overline{B}^0$ mixing should be considered.
The CP-violating asymmetries of $B^0(\overline{B}^0) \to {a_0}^+ {a_0}^-$ and ${a_0}^0 {a_0}^0$ decays are time dependent
and can be defined as
\beq
A_{\rm CP} &\equiv& \frac{\Gamma\left
(\overline{B}^0(\Delta t) \to f_{\rm CP}\right) -
\Gamma\left(B^0(\Delta t) \to f_{\rm CP}\right )}{ \Gamma\left
(\overline{B}^0(\Delta t) \to f_{\rm CP}\right ) + \Gamma\left
(B^0(\Delta t) \to f_{\rm CP}\right ) }\non
&=& A_{\rm CP}^{\rm dir} \cos(\Delta m  \Delta t) + A_{\rm CP}^{\rm mix} \sin (\Delta m \Delta t),
\label{eq:acp-def}
\eeq
where $\Delta m$ is the mass difference
between the two $B_d^0$ mass eigenstates, $\Delta t
=t_{\rm CP}-t_{tag} $ is the time difference between the tagged $B^0$
($\overline{B}^0$) and the accompanying $\overline{B}^0$ ($B^0$)
with opposite $b$ flavor decaying to the final CP-eigenstate
$f_{\rm CP}$ at the time $t_{\rm CP}$. The direct- and mixing-induced
CP-violating asymmetries $\acp^{\rm dir}  $
and $\acp^{\rm mix} $ can be written as
\beq
\acp^{\rm dir}  = \frac{ \left | \lambda_{\rm CP}\right |^2-1 } {1+|\lambda_{\rm CP}|^2},
\quad \acp^{\rm mix} = \frac{ 2 {\rm Im} (\lambda_{\rm CP})}{1+|\lambda_{\rm CP}|^2},
\label{eq:acp-dm}
\eeq
with the CP-violating parameter $\lambda_{\rm CP}$
\beq
\lambda_{\rm CP} &\equiv& \eta_f \;
\frac{V_{tb}^*V_{td}}{V_{tb}V_{td}^*} \cdot
\frac{ \langle f_{\rm CP} |H_{eff}|\overline{B}^0\rangle} {\langle f_{\rm CP} |H_{eff}|B^0\rangle}.
\label{eq:lambda2}
\eeq
where $\eta_f$ is the CP-eigenvalue of the final states.
Then the direct- and mixing-induced CP-violating
asymmetries for the $B^0 \to a_0^+ a_0^-$ and
$a_0^0 a_0^0$ decays in the pQCD approach
at LO level can be calculated
as,
\beq
\acp^{\rm dir}(B^0 \to a_0^+ a_0^-) &=&
31.0^{+3.7}_{-2.3}(\omega_{b})^{+10.4}_{-8.7}(B_{i}^{a_0})
^{+1.4}_{-1.4}({\rm CKM})   \%   \label{eq:adir98-pm}  \;,\\
\acp^{\rm mix}(B^0 \to a_0^+ a_0^-) &=&
0.9^{+9.2}_{-7.3}(\omega_{b})^{+7.4}_{-9.2}(B_{i}^{a_0})
^{+9.8}_{-9.6}({\rm CKM})   \%    \label{eq:amix98-pm}  \;,
\eeq
\beq
\acp^{\rm dir}(B^0 \to a_0^0 a_0^0) &=&
16.2^{+1.7}_{-1.1}(\omega_{b})^{+5.9}_{-4.9}(B_{i}^{a_0})
^{+0.7}_{-0.9}({\rm CKM})   \%   \label{eq:adir98-zz}  \;,\\
\acp^{\rm mix}(B^0 \to a_0^0 a_0^0) &=&
4.6^{+4.9}_{-3.6}(\omega_{b})^{+4.3}_{-4.8}(B_{i}^{a_0})
^{+9.9}_{-9.8}({\rm CKM})   \%    \label{eq:amix98-zz}  \;,
\eeq
where we have neglected the vanishing theoretical errors for the CP-violations in $B \to a_0 a_0$ decays arising from the
scalar decay constant $\bar f_{a_0}$ of $a_0$ meson.
It is interesting to see that these two channels,
namely, $B^0 \to a_0^+ a_0^-$ and $B^0 \to a_0^0 a_0^0$, have
large branching ratios and 
large direct CP asymmetries simultaneously, which could be easier to be measured at the running LHC experiments
and the forthcoming Super-B/Belle-II factory, and have the potential to reveal the QCD dynamics and the inner
structure involved in the light scalar $a_0$ meson.

\begin{figure}[!!hbt]
  \centering
  \begin{tabular}{ll}
  \includegraphics[height=5.0cm,width=6.0cm]{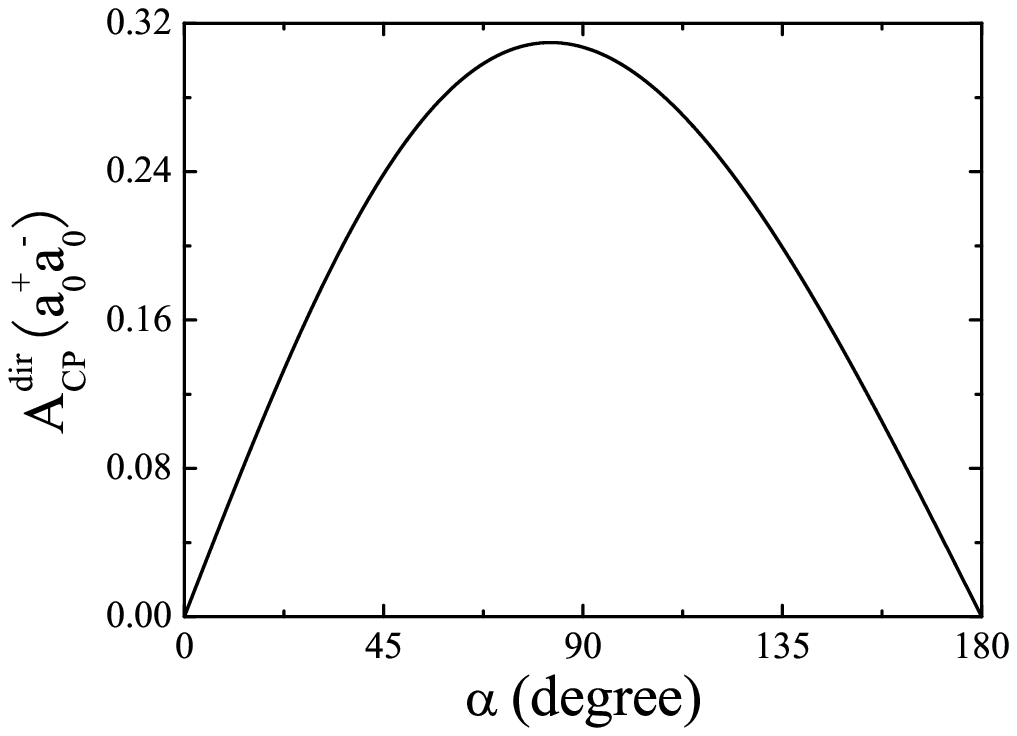}
 &\includegraphics[height=5.0cm,width=6.0cm]{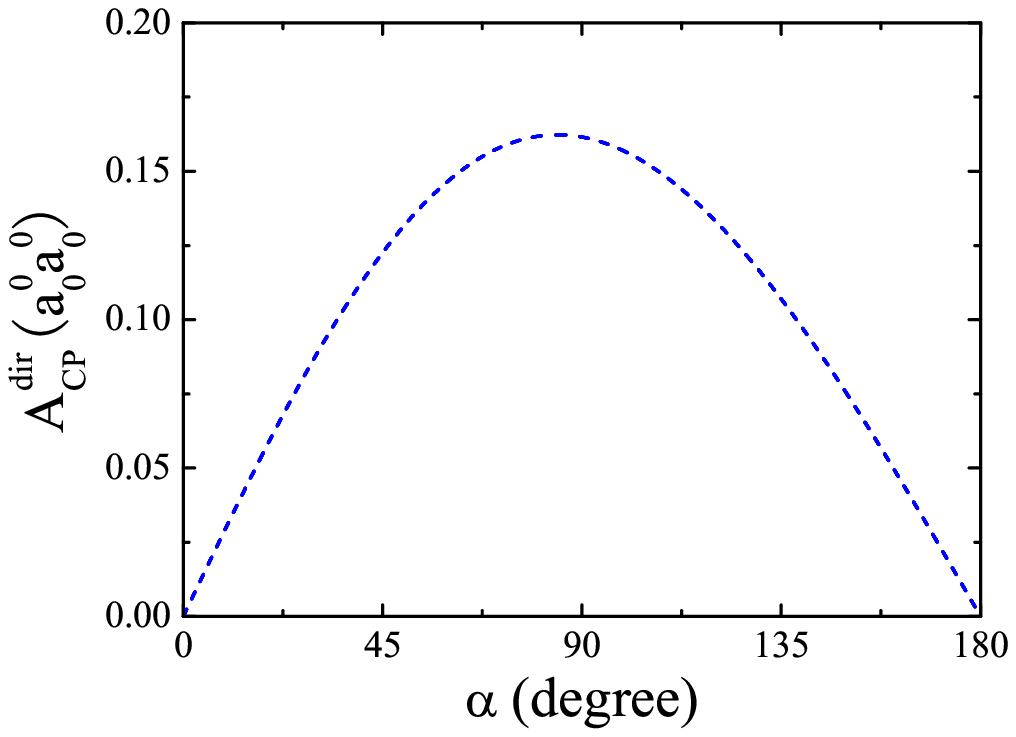}
  \end{tabular}
  \caption{(Color online) Dependence on the CKM angle $\alpha$ of
  the $B \to a_0^+ a_0^-$(Solid line) and $a_0^0 a_0^0$(Dashed line)
  direct CP violations at leading order in the pQCD approach,
  respectively.}
  \label{fig:fig3}
\end{figure}

Similarly, based on Eqs.~(\ref{eq:DA-B}),~(\ref{eq:DA-Bb}), and~(\ref{eq:adir98}), the direct CP-violating asymmetry can also be expressed as the function of the CKM angle $\alpha$,
\beq
A_{\rm CP}^{\rm dir}&=&\frac{2z\sin\alpha\sin\delta}
{1+2z\cos\alpha\cos\delta+z^2}\;.
\label{eq:acp-al}
\eeq
Then the precise measurements on these large direct CP violations
can also provide the constraints on the CKM angle $\alpha$
potentially. The variation of the direct CP-violating asymmetries
with the CKM angle $\alpha$ for the $B^0 \to a_0^+ a_0^-$(Solid line) and
$a_0^0 a_0^0$(Dashed line) decays is shown in Fig.~\ref{fig:fig3}. Again,
the central value about $90^\circ$ of the CKM angle $\alpha$
can be utilized to produce the above mentioned large direct
CP violations.

\section{Summary} \label{sec:summary}

In summary, we studied the two-body charmless hadronic $B \to a_0 a_0$
decays, which have the same Feynman diagrams as the $B \to \pi \pi$ modes at the quark level,
by employing the pQCD factorization approach based on the $k_T$ factorization theorem. Based on the assumption of
two-quark($q\bar q$) structure of the light scalar $a_0$ state,
we make 
theoretical predictions on the CP-averaged branching ratios
and the CP-violating asymmetries of the considered
$B \to a_0 a_0$ channels in the SM. Due to the large
non-factorizable contributions induced by the
asymmetric leading twist distribution amplitude of $a_0$ meson,
large branching ratios in the order of $10^{-6} \sim 10^{-5}$
have been predicted in the pQCD approach at LO level.
At the same time,
large direct CP violations around $15\%$ and $30\%$
in the $B^0 \to a_0^0 a_0^0$ and $a_0^+ a_0^-$
decays have also been observed. It is therefore
expected that the large branching ratios plus
the large CP asymmetries would be easier to be
measured at the running LHC experiments and
the forthcoming Super-B/Belle-II factory,
if $a_0$ is indeed the $q\bar q$ bound state.
Furthermore, the large non-factorizable
contributions in the $B \to a_0 a_0$ decays
can hint some important information on
resolving the famous $B \to \pi \pi$
puzzle, although this is
non-trivial work as clarified in the
literatures~\cite{Li:2009wba,Liu:2015sra}.
The detection of
these considered decays might be helpful to investigate the QCD dynamics in the channels and to
explore the inner structure of the light scalar $a_0$ state.
The investigation of 
the $B \to a_0 a_0$ decays could also provide more complementary constraints
on the CKM weak phase $\alpha$, since the same components as the $B \to \pi \pi$
modes exist in the considered $B \to a_0 a_0$ ones at the quark level. Frankly
speaking, the predictions in the present work suffered from large uncertainties
induced by the much less constrained hadronic parameters such as the Gegenbauer
moments $B_1^{a_0}$ and $B_3^{a_0}$, which need further studies in the
non-perturbative QCD(such as QCD sum rule and/or Lattice QCD) calculations
and the relevant experimental measurements(e.g., at BESIII, LHC, Super-B/Belle-II, etc.)
on the productions and/or decays involving the $a_0$ state.

\begin{acknowledgments}

This work is supported by the National Natural Science
Foundation of China under Grants No.~11205072, No.~11235005, and No.~11047014, and
by a project funded by the
Priority Academic Program Development
of Jiangsu Higher Education Institutions (PAPD).
\end{acknowledgments}


\begin{thebibliography}{99}
\expandafter\ifx\csname natexlab\endcsname\relax\def\natexlab#1{#1}\fi
\expandafter\ifx\csname bibnamefont\endcsname\relax
  \def\bibnamefont#1{#1}\fi
\expandafter\ifx\csname bibfnamefont\endcsname\relax
  \def\bibfnamefont#1{#1}\fi
\expandafter\ifx\csname citenamefont\endcsname\relax
  \def\citenamefont#1{#1}\fi
\expandafter\ifx\csname url\endcsname\relax
  \def\url#1{\texttt{#1}}\fi
\expandafter\ifx\csname urlprefix\endcsname\relax\def\urlprefix{URL }\fi
\providecommand{\bibinfo}[2]{#2}
\providecommand{\eprint}[2][]{\url{#2}}


\bibitem{Agashe:2014kda}
  K.~A.~Olive {\it et al.} [Particle Data Group Collaboration],
  Chin.\ Phys.\ C {\bf 38}, 090001 (2014).

\bibitem{Amhis:2014hma}
  Y.~Amhis {\it et al.} [Heavy Flavor Averaging Group Collaboration],
  arXiv:1412.7515; updated in
http://www.slac.stanford.edu/xorg/hfag.

\bibitem{Gershon:2006mt}
  T.~Gershon and A.~Soni,
  J.\ Phys.\ G {\bf 34}, 479 (2007).

\bibitem{Bona:2007qt}
  M.~Bona {\it et al.} [SuperB Collaboration],
  arXiv:0709.0451 [hep-ex];
  T.~Aushev {\it et al.} [Belle-II Collaboration],
  arXiv:1002.5012 [hep-ex].


\bibitem{Keum:2000ph}
  Y.~Y.~Keum, H.-n.~Li, and A.~I.~Sanda,
  Phys.\ Lett.\ B {\bf 504}, 6 (2001);
  Phys.\ Rev.\ D {\bf 63}, 054008 (2001).

\bibitem{Lu:2000em}
  C.~D.~L\"u, K.~Ukai, and M.~Z.~Yang,
  Phys.\ Rev.\ D {\bf 63}, 074009 (2001).


\bibitem{Li:2003yj}
  H.-n.~Li,
  Prog.\ Part.\ Nucl.\ Phys.\  {\bf 51}, 85 (2003).


\bibitem{Buchalla:1995vs}
  G.~Buchalla, A.J.~Buras, and M.E.~Lautenbacher,
  Rev.\ Mod.\ Phys.\  {\bf 68}, 1125 (1996).

\bibitem{Liu:2013lka}
  X.~Liu, Z.~J.~Xiao and Z.~T.~Zou,
  J.\ Phys.\ G {\bf 40}, 025002 (2013).

\bibitem{Cheng:2005nb}
  H.~Y.~Cheng, C.~K.~Chua, and K.~C.~Yang,
  Phys.\ Rev.\ D {\bf 73}, 014017 (2006);
  {\it ibid.} {\bf 77}, 014034 (2008).

\bibitem{Cheng:2009xz}
  H.~Y.~Cheng and J.~G.~Smith,
  Ann.\ Rev.\ Nucl.\ Part.\ Sci.\  {\bf 59}, 215 (2009).

\bibitem{Liu:2015sra}
  X.~Liu, H.-n.~Li and Z.~J.~Xiao,
  Phys.\ Rev.\ D {\bf 91}, 114019 (2015).

\bibitem{Bauer:1986bm}
  M.~Bauer, B.~Stech and M.~Wirbel,
  Z.\ Phys.\ C {\bf 34}, 103 (1987);
  M.~Wirbel, B.~Stech and M.~Bauer,
  Z.\ Phys.\ C {\bf 29}, 637 (1985).

\bibitem{Diehl:2001xe}
  M.~Diehl and G.~Hiller,
  \jhep {\bf 06}, 067 (2001);
  S.~Laplace and V.~Shelkov,
  Eur.\ Phys.\ J.\ C {\bf 22}, 431 (2001).

\bibitem{Beneke:1999br}
  M.~Beneke, G.~Buchalla, M.~Neubert and C.~T.~Sachrajda,
  Phys.\ Rev.\ Lett.\  {\bf 83}, 1914 (1999);
  Nucl.\ Phys.\ B {\bf 591}, 313 (2000).

\bibitem{Du:2000ff}
  D.~s.~Du, D.~s.~Yang and G.~h.~Zhu,
  Phys.\ Lett.\ B {\bf 488}, 46 (2000).

\bibitem{Li:1996gi}
  H.-n.~Li,
  Phys.\ Rev.\ D {\bf 55}, 105 (1997);
  Phys.\ Lett.\ B {\bf 405}, 347 (1997).

\bibitem{Bauer:2004tj}
  C.~W.~Bauer, D.~Pirjol, I.~Z.~Rothstein and I.~W.~Stewart,
  Phys.\ Rev.\ D {\bf 70}, 054015 (2004).


\bibitem{Arnesen:2006vb}
  C.M.~Arnesen, Z.~Ligeti, I.Z.~Rothstein, and I.W.~Stewart,
  Phys.\ Rev.\ D {\bf 77}, 054006 (2008).

\bibitem{Chay:2007ep}
  J.~Chay, H.-n.~Li, and S.~Mishima,
  Phys.\ Rev.\ D {\bf 78}, 034037 (2008).

\bibitem{Lu:2002iv}
  C.D.~L\"u and K.~Ukai,
  Eur.\ Phys.\ J.\ C {\bf 28}, 305 (2003).

\bibitem{Li:2004ep}
  Y.~Li, C.D.~L\"u, Z.J.~Xiao, and X.Q.~Yu,
  Phys.\ Rev.\ D {\bf 70}, 034009 (2004).

\bibitem{Ali:2007ff}
  A.~Ali, G.~Kramer, Y.~Li, C.D.~L\"u, Y.L.~Shen, W.~Wang, and Y.M.~Wang,
  Phys.\ Rev.\ D {\bf 76}, 074018 (2007).


\bibitem{Xiao:2011tx}
  Z.J.~Xiao, W.F.~Wang, and Y.Y.~Fan,
  Phys.\ Rev.\ D {\bf 85}, 094003 (2012);
  Y.L.~Zhang,  X.Y.~Liu, Y.Y.~Fan, S.~Cheng, and Z.J.~Xiao, Phys.\ Rev.\ D {\bf 90}, 014029 (2014).

\bibitem{Li:2005kt}
  H.-n.~Li, S.~Mishima, and A.~I.~Sanda,
  Phys.\ Rev.\ D {\bf 72}, 114005 (2005).

\bibitem{Lu:2002ny}
  C.~D.~L\"u and M.~Z.~Yang,
  Eur.\ Phys.\ J.\ C {\bf 28}, 515 (2003).

\bibitem{Li:2008tk}
  R.~H.~Li, C.~D.~L\"u, W.~Wang and X.~X.~Wang,
  Phys.\ Rev.\ D {\bf 79}, 014013 (2009).

\bibitem{Aubert:2004hs}
  B.~Aubert {\it et al.} [BaBar Collaboration],
  Phys.\ Rev.\ D {\bf 70}, 111102 (2004);
  {\it ibid.} {\bf 75}, 111102 (2007);
  S.~Uehara {\it et al.} [Belle Collaboration],
  Phys.\ Rev.\ D {\bf 80}, 032001 (2009);
  R.~Aaij {\it et al.} [LHCb Collaboration],
  Phys.\ Rev.\ D {\bf 88}, 072005 (2013).


\bibitem{Chernyak:1983ej}
  V.~L.~Chernyak and A.~R.~Zhitnitsky,
  Phys.\ Rept.\  {\bf 112}, 173 (1984);
A.~R.~Zhitnitsky, I.~R.~Zhitnitsky and V.~L.~Chernyak,
  Sov.\ J.\ Nucl.\ Phys.\  {\bf 41}, 284 (1985)
  [Yad.\ Fiz.\  {\bf 41}, 445 (1985)];
  V.~M.~Braun and I.~E.~Filyanov,
  Z.\ Phys.\ C {\bf 44}, 157 (1989)
  [Sov.\ J.\ Nucl.\ Phys.\  {\bf 50}, 511 (1989)]
  [Yad.\ Fiz.\  {\bf 50}, 818 (1989)];
  V.~M.~Braun and I.~E.~Filyanov,
  Z.\ Phys.\ C {\bf 48}, 239 (1990)
  [Sov.\ J.\ Nucl.\ Phys.\  {\bf 52}, 126 (1990)]
  [Yad.\ Fiz.\  {\bf 52}, 199 (1990)].


\bibitem{Ball:1998tj}
  P.~Ball,
  \jhep {\bf 09}, 005 (1998);
  P.~Ball,
  \jhep {\bf 01}, 010 (1999).

\bibitem{Li:2009wba}
  H.-n.~Li and S.~Mishima,
  Phys.\ Rev.\ D {\bf 83}, 034023 (2011);
 {\it ibid.} {\bf 90}, 074018 (2014).





\end{thebibliography}
\end{document}